\documentclass[preprint,12pt]{elsarticle}
\usepackage{epsf,epsfig}
\usepackage[psamsfonts]{amssymb}
\usepackage{amsmath}
\usepackage{bm}
\usepackage{graphicx}
\usepackage{graphicx,epstopdf}
\usepackage{caption}
\usepackage{subcaption}

\begin{document}

\begin{frontmatter}
\title{Scattering of nonlinear coherent state on sphere by an absorptive and dispersive dielectric slab}
\author{ Roghayeh Asadi Aghbolaghi$^1$, Ehsan Amooghorban\footnote{Ehsan.amooghorban@sci.sku.ac.ir}$^{1,2}$, and Ali Mahdifar$^{1,2}$}

\address{$^1$  Department of Physics, Faculty of Basic Sciences, Shahrekord University, P.O. Box 115, Shahrekord 88186-34141, Iran.}
\address{$^2$  Photonic Research Group, Shahrekord University, P.O. Box 115, Shahrekord 88186-34141, Iran.}

\begin{abstract}
In this paper, we examine the modification of specific nonclassical properties of the nonlinear coherent state on sphere upon perpendicular propagation through an absorptive and dispersive dielectric slab at finite temperature. For this purpose, by describing the dielectric dispersion of the slab by Lorentz model, the quadrature squeezing and the Mandel parameter are evaluated for the transmitted state.
A generalization of single-mode nonlinear coherent state to the continuum-mode is considered.
The degree of second-order coherence is instead calculated for a continuum nonlinear coherent state on sphere, and the quantum noise effects produced by transmission through the slab on the antibunching feature are examined.
We find that near the medium resonance the detrimental effect
of the loss and thermal fluctuations of the slab are not compensate with increasing the physical space curvature of the incident state.
\end{abstract}


\end{frontmatter}

\section{Introduction}\label{Sec:Introduction}
It is well known that optical devices such as mirrors, beam splitters, interferometers, optical fibers,
and detectors composed of optical components of finite extent made from lossy media.  The material properties of these components play an important role in the outcome of experiments.  Therefore, the use of such optical instruments needs careful examination of their action on the optical properties of the radiation field.

In classical electrodynamic, the material properties of homogenous and isotropic dielectric media are macroscopically described by electric permittivity $\varepsilon(\omega)$.
The electric permittivity of a dispersive and absorbing dielectric medium is expressed as a complex function of frequency
which its imaginary and real parts are related to each other by Kramers -Kronig relations.
Since the imaginary part of dielectric function is associated with loss in the dielectric medium,
therefore, there is inevitably noise due to the fluctuation-dissipation theorem~\cite{Milonni1994}. Thus, absorption in dielectric media add noise to a beam of light.
The optical properties of the light that propagate through the bounded medium will be modified by absorption,
dispersion, quantum noises, and the various reflections which take place from the medium surfaces.
Some of these modifications gives rise to identical effects in both classical and quantum domain.

The problem of describing the influence of dispersive and absorptive dielectric media on classical optical pulses
has been studied in detail in the literature~\cite{Garrett1970,Halevi1986}. However, there are some modifications
on the incident light with nonclassical nature that can only be described in the framework of a full quantum theory.
In Refs.~\cite{Artoni1997,Artoni1998b,Matloob2000}, by applying a phenomenological approach to quantize the electromagnetic field
in absorptive and dispersive dielectric media and evaluating the Poynting vector,
it has been shown that the incident quantum optical pulse deforms and its optical properties change due to the loss.
A similar treatment has been employed to the case of an amplifying slab~\cite{Artoni1998a}.

Recently, a canonical formalism is used to study the
effective-medium theories in quantum optics via the propagation
of the squeezed light through a loss-compensated
metamaterial~\cite{Amooghorban2013}, and as well for propagation
in arbitrary directions in layered
metamaterials~\cite{Amooghorbanarxive}. For active plasmonics and
for some passive metamaterials, it is shown that an additional
effective-medium parameter is indispensable besides the effective
index, namely the effective noise-photon distribution. Also, it
has been successfully applied to the study of input-output
relation for anisotropic magnetodielectric multilayer
structures~\cite{Hoseinzadeh2017a}, the spontaneous
emission~\cite{Kheirandish2010,Behbahani2016}, nonlinear
magnetodielectric media~\cite{Amooshahi2010,Kheirandish2011},
moving media~\cite{Hoseinzadeh2017b}, and the Casimir
effect~\cite{Amooghorban2011}. Since nonclassical states are an
important prerequisite for different tasks of quantum information
processing, such as quantum cryptography, quantum computing,
quantum voting~\cite{Kilin2014}, in the present paper, we utilize
the above-mentioned rigorous canonical approach to study the
behavior of a nonclassical state (nonlinear coherent state on a
sphere) in passing through a dielectric slab. From this point of
view, nonlinear coherent states which possess prominent
nonclassical properties are good candidates for examination by
transmission through a dielectric medium.

Coherent states (CSs) of the harmonic oscillator \cite{star} as
well as generalized CSs associated with various algebras
\cite{dstar,tav kla,tav ali(2000)} play an important role in
various fields of physics and in particular, they have found
considerable applications in quantum optics. Among the
generalized CSs, the so-called nonlinear CSs or f-deformed CSs
\cite{star3} have attracted much attention in recent years,
mostly because they exhibit nonclassical properties, such as
amplitude squeezing and quantum interference
\cite{dstar3,cstar3}. These states, which could be realized in
the center of mass motion of an appropriately laser-driven
trapped ion \cite{cstar3}, \cite{Vogel} and in a micromaser under
intensity-dependent atom-field interaction \cite{chstar3}, are
associated with nonlinear algebras and defined as the eigenstates
of the annihilation operator of a f-deformed oscillator
\cite{star3,tav manko}. In Ref. \cite{Ali 1}, the two dimensional
harmonic oscillator on a curved space (sphere) has been studied.
It has been found that a two-dimensional harmonic oscillator
algebra may be considered as a deformed one-dimensional harmonic
oscillator algebra. Moreover, the algebra of an oscillator on a
sphere represents a deformed version of the oscillator algebra in
flat space where the curvature of the physical space plays the
role of deformation parameter. For the sphere coherent states
(SCSs), which are special realizations of the nonlinear CSs, the
influence of the curvature of the space on the algebraic
structure and the nonclassical properties have been analyzed.
Also, by using the SCSs, the curvature
effects on some physical phenomena have been studied in Refs. \cite{Ali 3,Josa,BS,Amooghorban 2015A}.

Recently in Ref.~\cite{Amooghorban2014}, the influence of a
dielectric slab at finite temperature on the Wigner function and
the entanglement degree of outgoing states have been investigated
for the case that the input state is the nonlinear CS on a sphere.
The results indicated two competing trends arise: the loss and
thermal effects degrade the specific quantum properties of these
states, such as coherence and entanglement. Whereas, the
rise of the curvature parameter leads to enhance the nonclassical
properties of the output state when the incident light is tuned
far from the medium resonance.
%

With the above background, a question still remains: what happens to quantum-statistical properties of the transmitted nonlinear CS on a sphere through the slab at finite temperature? Does the nonclassical properties of the incident state such as antibunching, sub-Poissonian statistics and squeezing, survive after propagation?
%
In order to answer these questions, we evaluate the quadrature squeezing, the Mandel parameter and the degree of the second-order
correlation function of the outgoing state from the slab. Finally, the competition between various effects such as the quantum noises and the physical curvature of the incident state are examined.
It would be seen that the detrimental effect of loss on the incident light can not compensate even when the curvature parameter of the incident state increased.

The layout of the paper is as follows: In Sec.~\ref{Sec:Basic
equations}, the quantization scheme is briefly summarized and the
input-output relation for the electromagnetic field operators at
the boundaries of the dielectric slab is introduced. In
Sec.~\ref{Sec:Nonlinear coherent states on sphere}, the nonlinear
CSs on a sphere will be introduced. Subsequently, in
Secs.~\ref{Sec:Quadrature Squeezing} and~\ref{Sec:Mandel
parameter}, we calculate the quadrature squeezing and Mandel
parameter for quantum states outgoing of the dielectric slab when
the incident states from right and left sides are, respectively,
the nonlinear SCS and vacuum. Here, the dissipative and dispersive
effects of the dielectric slab enter into our calculations using
well known Lorentz model.
In Sec.~\ref{Sec:The degree of second order quantum coherence}, a
generalization of single-mode nonlinear SCS to the continuum-mode
one is carried out. Then, by calculating the second order quantum
coherence function, the antibunching property of the scattered
continuum-mode nonlinear SCS by the dielectric slab are studied.
Finally, conclusions are given in Sec.~\ref{Sec:Conclusion}.

\section{Basic equations}\label{Sec:Basic equations}
%
In this section, we briefly study the quantum formalism of scattering of the electromagnetic waves from an absorptive and dissipative dielectric slab
with thickness $2l$. The permittivity function of the slab is given by $\varepsilon \left( \omega  \right)$
and we assume to be an arbitrary complex function of frequency, satisfying Kramers - Kronig relations.
The refractive index of the entire system, $n\left( \omega  \right)$, is defined in terms of the dielectric permittivity function, $\varepsilon \left( \omega  \right)$, as follows,
\begin{eqnarray}\label{refractive index}
n(\omega)=\left\{ {\begin{array}{*{20}c}
\eta (\omega)+i\,\kappa (\omega)=\sqrt{\varepsilon( \omega)}\,\,\,\,\,\,for\,\,\,\,\left| x \right|\le l ,\\
1\,\,\,\,\,\,\,\,\,\,\,\,\,\,\,\,\,\,\,
\,\,\,\,\,\,\,\,\,\,\,\,\,\,\,\,\,\,\,\,\,\,\,\,\,\,\,\,\,\,\,\,\,\,\,\,\,\,\,\,\,\,\,for\,\,\,\,\,\left| x \right|> l ,\\
\end{array}} \right.
\end{eqnarray}
where $\eta (\omega )$ and $\kappa(\omega )$ are real and imaginary parts of refractive index, respectively.

\begin{figure}[t]
\includegraphics[width=6.5cm]{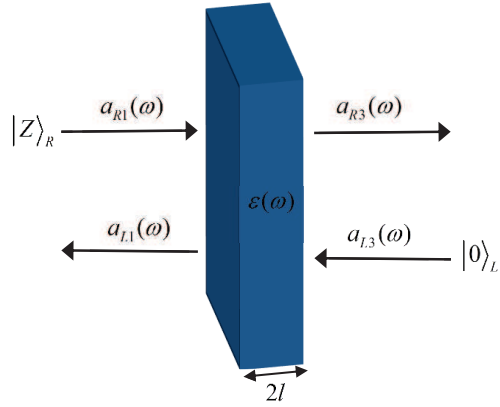}
\centering
\caption{Representation of the operators of the input and output modes defined in the relation~(\ref{scattering matrix}). Here, the states of the input modes impinging leftwards and rightwards on the slab have been shown by $\left|\  \right\rangle_{R} $ and $\left| \ \right\rangle_{L} $.}
\label{Fig:slab}
\end{figure}

Following the canonical approach in quantization of the electromagnetic field in the presence of dissipative, and dispersive dielectric medium. We start from an appropriate Lagrangian describing the electromagnetic field, the medium and their interaction between them. It can be accomplished consistently by modeling the medium by a reservoir composed of a continuum of three dimensional harmonic oscillators~\cite{Hopfield 1958}.
These harmonic oscillators describe both polarizability and the absorption of the medium and interact with the electric field via a dipole interaction
term. In this way, the medium explicitly introduces into the formalism. 
By defining the canonical conjugate momentums of system and imposing commutation relations between the dynamic variables and their conjugates, the Hamiltonian of the whole system emerging from the canonical procedure. The constitutive equations of the medium are obtained as the consequences of the Heisenberg equations of the system and the dielectric permeability of the medium is calculated in terms of the parameters applied in the theory.
Combining these Heisenberg equations lead to a Langevin-Schrödinger equation for the
vector potential operator, wherein, the explicit form of the noise current density is known.
This Langevin equation is considered as the
basis of the macroscopic description of the electromagnetic field coupled to the medium in phenomenological approach~\cite{Artoni1997,Artoni1998b,Gruner 1996,Dung 1998,Scheel 1998,Matloob 1995,Knoll 2001,Dung 2000,Dung 2003,Matloob 2004}.
More details concerning this rigorous canonical approach can be found in Refs.~\cite{Amooghorban2013,Amooghorbanarxive,Hoseinzadeh2017a,Kheirandish2010,Huttner1992,Jeffers1996,Suttorp2004,Amooshahi 2009,Philbin 2010}.
According to the canonical formalism, the positive frequency component of the vector potential operator for the case that the radiation propagates in $x$ direction, perpendicular to the slab, is written as~\cite{Amooghorban2013,Amooghorbanarxive,Hoseinzadeh2017a}
\begin{eqnarray}\label{vector potential}
\hat{A}^{+}_{\Omega}(x,t)&=&\int_{0}^{+\infty} d\omega \left(\frac{\hbar}{4\pi\varepsilon_{0}c\omega\sigma}\right)^{1/2}\left[{\hat{a}_{R\Omega}(\omega) e^{-i\omega x/c}}\right.\left.{+\hat{a}_{L\Omega}(\omega) e^{i\omega x/c}}\right]  e^{-i\omega t},\,\,\,\,\,\,\,\,
\end{eqnarray}
where $\Omega =1,3$ indicates left and right sides of the slab, respectively, $\sigma$ is the area of quantization in $yz$ plane and the indices $R$ and $L$ refer to the rightward and leftward propagating modes on the dielectric slab.
In Fig.~\ref{Fig:slab}, a schema of the geometry of the electromagnetic waves scattered by dielectric slab has been shown in terms of the annihilation operators defined in Eq.~(\ref{vector potential}).
%

%
In the presence of absorbing dielectric slab, the operators of the output modes, ${{\hat{a}}_{R1}}$ and ${{\hat{a}}_{L3}}$, by means of input-output relation can be expressed in term of the operators of the input modes, ${\hat{a}}_{L1}$ and ${\hat{a}}_{R3}$, and the noise operators, $\hat{F}_{L}$ and $\hat{F}_{R}$, as follows~\cite{Amooghorban2013,Amooghorbanarxive}
\begin{eqnarray}\label{scattering matrix}
\left( {\begin{array}{*{20}c}
   {{{\hat{a}}}_{L1}}( \omega) \\
  {{{\hat{a}}}_{R3}}( \omega) \\
\end{array} }\right)&=&\left( {\begin{array}{*{20}c}
   R\left( \omega  \right)\,\,\,\,\,T( \omega) \\
  T\left( \omega  \right)\,\,\,\,\,\,R( \omega) \\
\end{array}} \right)\left( {\begin{array}{*{20}c}
   {{{\hat{a}}}_{R1}}( \omega) \\
  {{{\hat{a}}}_{L3}}( \omega) \\
\end{array}} \right)+\left( {\begin{array}{*{20}c}
   {{{\hat{F}}}_{L}}( \omega) \\
  {{{\hat{F}}}_{R}}( \omega) \\
\end{array} }\right),
\end{eqnarray}
where
\begin{eqnarray}\label{R coefficient}
R ( \omega )=\frac{\left( {{n}^{2}}\left( \omega  \right)-1 \right)\exp \left[ {-2i\omega l}/{c}\; \right]\left[ \exp \left( {4i\omega n( \omega )l}/{c}\; \right)-1 \right]}{{{\left( n( \omega )+1 \right)}^{2}}-{{\left( n( \omega )-1 \right)}^{2}}\exp \left[ {4i\omega n( \omega )l}/{c}\; \right]},
\end{eqnarray}
and
\begin{equation}\label{T coefficient}
T(\omega )=\frac{4n\left( \omega  \right)\exp \left[ {2i\omega \left( n\left( \omega  \right)-1 \right)l}/{c}\; \right]}{{{\left( n\left( \omega  \right)+1 \right)}^{2}}-{{\left( n\left( \omega  \right)-1 \right)}^{2}}\exp \left[ {4i\omega n\left( \omega  \right)l}/{c}\; \right]},
\end{equation}
are the amplitudes of reflection and transmission coefficients, respectively. The outward-propagating noise operator ${{\hat{F}}_{L}}$, arising from the dissipative nature of the dielectric slab, is defined as
\begin{eqnarray}\label{F operator}
\hat{F}_{L}(\omega)&=&i\sqrt{\frac{2\omega \eta(\omega)\kappa(\omega)}{c}}\\
&&\times\int_{-l}^{+l}{d{x}'} \left[ V(\omega)e^{-i\omega n(\omega){x}'/c}\right.\left.+W(\omega)e^{+i\omega n(\omega){x}'/c}\right] \hat{f}({x}',\omega ),\nonumber
\end{eqnarray}
where the coefficients $V( \omega )$  and $W( \omega )$ in the integrand of the noise operator are
\begin{eqnarray}
V(\omega)=\frac{2\left( n( \omega )+1 \right)\exp \left[ {i\omega \left( n( \omega )-1 \right)l}/{c}\; \right]}{{{\left( n\left( \omega  \right)+1 \right)}^{2}}-{{\left( n\left( \omega  \right)-1 \right)}^{2}}\exp \left[ {4i\omega n\left( \omega  \right)l}/{c}\; \right]},\,\,\,\,\,\,\,\, \\
W(\omega)=\frac{2\left( n\left( \omega  \right)-1 \right)\exp \left[ {i\omega \left( 3n\left( \omega  \right)-1 \right)l}/{c}\; \right]}{{{\left( n\left( \omega  \right)+1 \right)}^{2}}-{{\left( n\left( \omega  \right)-1 \right)}^{2}}\exp \left[ {4i\omega n\left( \omega  \right)l}/{c}\; \right]}.\,\,\,\,\,\,\,\,
\end{eqnarray}
The form of the noise operator on the right, $\hat{F}_{R}(\omega)$, is obtained from that on the left by the simple prescription $x'\rightarrow -x'$.
Here, the basic bosonic operator $\hat{f}( x,\omega )$, associated
with the electric excitation within slab, having the commutation properties of the form
\begin{equation}\label{commutation relation f}
\left[\hat{f}( x,\omega  ),{\hat{f}^{\dagger }}( x',{\omega}') \right]=\delta \left( x-x' \right)\delta\left( \omega -{\omega}' \right).
\end{equation}
By using the commutator~(\ref{commutation relation f}) and the forms of the
coefficients~(\ref{R coefficient}) and~(\ref{T coefficient}), it is straightforward to show that the noise operators $\hat{F}_\Lambda$ and $\hat{F}_\Lambda^{\dagger}$ satisfy the following commutation relations
\begin{equation}\label{commutation relation F}
\left[\hat{F}_\Lambda\left( \omega  \right),{\hat{{F}}_\Lambda^{\dagger }}({\omega }') \right]=\left( 1-{{\left| R\left( \omega  \right) \right|}^{2}}-{{\left| T\left( \omega  \right) \right|}^{2}} \right)\delta \left( \omega -{\omega }' \right),
\end{equation}
where $\Lambda=L,R$. It follows from the above expressions that the input and output mode operators in the input-output relation~(\ref{scattering matrix}) have the following commutation relations
\begin{equation}\label{commutation relation aR1 and aR1dagger}
\left[ {\hat{{a}}_{R1}}\left( \omega  \right), \hat{a}_{R1}^{\dagger }\left( {{\omega }'} \right) \right]=\left[ {\hat{{a}}_{L3}}\left( \omega  \right),\hat{a}_{L3}^{\dagger }\left( {{\omega }'} \right) \right]=\delta \left( \omega -{\omega }' \right),
\end{equation}
\begin{equation}\label{commutation relation aR1 and aL3}
\left[ {\hat{{a}}_{R1}}\left( \omega  \right), \hat{a}_{L3}^{\dagger }\left( {{\omega }'} \right) \right]=\left[ {\hat{{a}}_{L3}}\left( \omega  \right),\hat{a}_{R1}^{\dagger }\left( {{\omega }'} \right) \right]=0,
\end{equation}
\begin{equation}\label{commutation relation aL1 and aL1dagger}
\left[ {\hat{{a}}_{L1}}\left( \omega  \right),\hat{a}_{L1}^{\dagger }\left( {{\omega }'} \right) \right]=\left[ {\hat{{a}}_{R3}}\left( \omega  \right),\hat{a}_{R3}^{\dagger }\left( {{\omega }'} \right) \right]=\delta \left( \omega -{\omega }' \right).
\end{equation}
The foregoing relations provide us with a theoretical tool to
investigate the effects of perpendicular propagation through the slab on the incident nonlinear SCSs.
%

\section{ Nonlinear coherent states on a sphere}\label{Sec:Nonlinear coherent states on sphere}
%
In Ref.~\cite{Ali 1}, we searched for a relation between the deformation
function of the f-deformed oscillator algebra and a two-dimensional harmonic
oscillator on sphere. We found that we could
consider a two-dimensional harmonic oscillator algebra as a deformed
one-dimensional harmonic oscillator algebra
\begin{equation}\label{tav 3}
[\hat{A},\hat{A}^{\dag}]=(\hat{n}+1)f^{2}(\hat{n}+1)
-\hat{n}f^{2}(\hat{n}),
\end{equation}
where
\begin{eqnarray}\label{tav 1}
\hspace{3cm}\hat{A}&=&\hat{a}f(\hat{n})=f(\hat{n}+1)\hat{a},\\
\hspace{3cm}\hat{A}^{\dag}&=&f^{\dag}(\hat{n})\hat{a}^{\dag}=\hat{a}^{\dag}f^{\dag}(\hat{n}+1).
\end{eqnarray}
We obtained the deformation function
corresponding to sphere harmonic oscillators as
\begin{equation}\label{ali 6}
f_{s}(n)=f_{f}(\hat{n})g(\lambda,n),
\end{equation}
where the deformation function
corresponding to flat harmonic oscillators, $f_{f}(n)$, is defined as
\begin{equation}\label{ali 5}
f_{f}(\hat{n})=\sqrt{(N+1-\hat{n})},
\end{equation}
and
\begin{eqnarray}\label{g definition}
g\left( \lambda ,n \right) &=&
\sqrt{\left( \lambda \left( N+1-n \right)+\sqrt{1+{{{\lambda }^{2}}}/{4}}\right) \left( \lambda n+\sqrt{1+{{{\lambda }^{2}}}/{4}\;} \right)}.
\end{eqnarray}
The parameter $\lambda=\frac{1}{R^{2}}$ is the curvature of sphere
and $N+1$ is the dimension of the associated Fock space.
It is obvious that in the flat limit, $g(\lambda,n)\rightarrow 1$
and Eq. (\ref{ali 6}) reduces to (\ref{ali 5}).
Therefore, we can consider the two-dimensional harmonic
oscillator algebra as a type of one-dimensional deformed
harmonic oscillator with the deformation function $f(n)$ and also the sphere
oscillator algebra as a deformed version of the flat oscillator algebra with the deformation
function $g(\lambda,n)$.
Now, by defining the relation
\begin{equation}\label{}
\hat{A}|0\rangle=0=\hat{A}^{\dag}|N\rangle,
\end{equation}
for each constant value of $N$, we encounter with a
$N$-dimensional Hilbert space. Therefore, the constructed SCS is
\begin{eqnarray}\label{Nonlinear coherent state}
\left| Z \right\rangle
=\frac{1}{\sqrt{M}}\exp(\mu\hat{A}^{\dag})|0\rangle= \frac{1}{\sqrt{M}}\sum\limits_{n=0}^{N}{\sqrt{\left( \begin{array}{*{20}c}
N \\
n \\
\end{array} \right)}\left[ g\left( \lambda ,n \right)\right]}! \,{{Z}^{n}}\left| n \right\rangle ,
\end{eqnarray}
where by definition $\left[ g\left( \lambda ,0 \right)\right]!=1$ and
\begin{equation}\label{}
\left[ g\left( \lambda ,n \right)\right]!=g\left( \lambda ,n \right)g\left( \lambda ,n-1 \right)\cdots g\left( \lambda ,1 \right),
\end{equation}
$Z$ is a complex number and the normalization constant $M$ is
defined by
$M=\sum\limits_{n=0}^{N}{{\left( \begin{array}{*{20}c}
 N \\
 n \\
\end{array} \right)}}{{\left( \left[ g\left( \lambda ,n \right) \right]! \right)}^{2}}{{\left| Z \right|}^{2n}}$.

Some quantum statistical properties of such constructed nonlinear SCSs, including mean number of photons,
Mandel parameter and quadrature squeezing, were theoretically examined in~\cite{Ali 1}. The results show that
the curvature of physical space leads to the enhancement of
nonclassical properties of the SCSs.

Furthermore, a realizable physical model to generate the SCSs has
been proposed in Ref.~\cite{Vogel}. Based on the nonlinear CSs of
the atomic center-of-mass motion of a trapped ion, one may control
the laser-ion interaction in such a way that nonlinear SCSs are
obtained as motional dark states of the system. Furthermore, the
Rabi frequencies may be properly adjusted to simulate and vary
the curvature of the space in which the nonlinear CSs under study
are prepared.

In this paper, we intend to examine how nonclassical properties of the nonlinear SCS on a sphere become modified after propagation through an absorptive and dispersive dielectric slab. To do this, we first need
to prepare a state of radiation filed which can be described by
the nonlinear SCS on a sphere. In Ref.~\cite{BS}, a scheme for the
generation of an arbitrary SCS has been proposed in a single-mode resonator by the
transfer of atomic coherence to the cavity field. In this model,
the atomic system consists of a series of two-level atoms,
initially prepared in a $\lambda$-dependent superposition of the
exited state and the ground state. These atoms interact with a
resonant mode of the electromagnetic field in cavity which is
initially in the vacuum state via the Jaynes-Cummings Hamiltonian.

\begin{figure}[t]
\includegraphics[width=7.5cm]{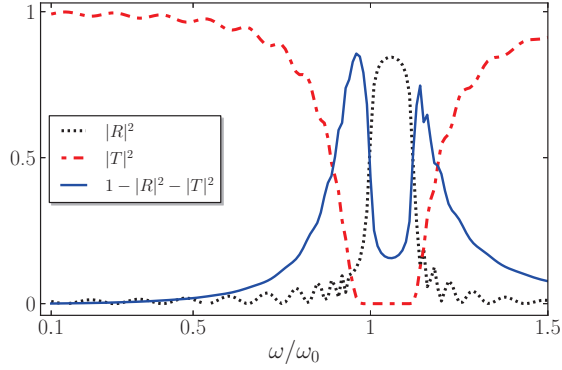}
\centering
\caption{The reflection coefficient $|R|^2$ (solid blue curve), the transmission coefficient $|T|^2$ (dashed red curve)
and the absorption coefficient $1-|R|^2-|T|^2$ (dotted gray curve) of a dielectric slab with thickness $l=10 c/{\omega }_{0}$
are shown as a function of ${{{\omega }_{\,}}}/{{{\omega }_{\,0}}}$. The parameters are chosen in Lorentz model~(\ref{Lorentz model})
are ${{{\omega }_{p\,}}}/{{{\omega }_{\,0}}}\;=0.5$ and ${\gamma }/{{{\omega }_{\,0}}}\;=0.01$.}
\label{Fig:R2 and T2 versus omega}
\end{figure}

\section{Quadrature Squeezing}\label{Sec:Quadrature Squeezing}
%

In order to investigate the dissipative and dispersive effects of the dielectric slab on the incident quantum state, the expectation value of the physical variable of our system is needed. In this and next section, we consider the general state of the system as
\begin{equation}\label{general state of the system}
\left| \varphi  \right\rangle ={{\left| Z \right\rangle }_{R}}{{\left| 0 \right\rangle }_{L}}\left| F \right\rangle.
\end{equation}
Here, the quantized field impinging from the right and the left
are taken as the conventional vacuum, ${{\left| 0 \right\rangle
}_{L}}$, and the nonlinear SCS on a sphere, ${{\left| Z
\right\rangle }_{R}}$, respectively, and the noise state of the
slab is represented by $\left| F
\right\rangle $. Now, in order to study the noise properties of
the transmitted light, we calculate the variance of the
quadrature field operators which associated with the outgoing
field in domain 3. The quadrature field operators of the
transmitted field in domain 3 are defined as~\cite{Scully1997}
\begin{eqnarray}\label{the quadrature operators}
{{\hat{X}}_{R3}}&=&\frac{1}{2}\left( {{{\hat{a}}}_{R3}}+\hat{a}_{R3}^{\dagger } \right),\\
{{\hat{Y}}_{R3}}&=&\frac{1}{2i}\left( {{{\hat{a}}}_{R3}}-\hat{a}_{R3}^{\dagger } \right),
\end{eqnarray}

By using the usual definition of the root-mean-square deviation, after some lengthy calculations, the uncertainty in the quadrature operator ${{\hat{X}}_{R3}}$ is given by
\begin{eqnarray}\label{uncertainty quadrature X}
&&\langle\varphi|(\Delta\hat{X}_{R3})^{2}|\varphi\rangle=\langle\varphi|\hat{X}_{R3}^{2}|\varphi\rangle
-\langle\varphi|\hat{X}_{R3}|\varphi\rangle^{2}\nonumber\\
&&=\frac{1}{4M}\sum\limits_{n =0}^{N}\sqrt{\binom{N}{n}}[g(\lambda, n)!]\nonumber\\
&&\times\left\{\left({\sqrt{\binom{N}{n}}[g(\lambda, n)!] |Z|^{2n} (2n|T|^{2}+1+2\langle F| F^{\dag}_{R}F_{R}|F\rangle)}\right.\right.\nonumber\\
&&\hspace{0.4cm}\left.{+\sqrt{\binom{N}{n+2}}[g(\lambda, n+2)!]\sqrt{(n+1)(n+2)}(Z^{2}T^{2}+Z^{\ast2}T^{\ast2})}\right)\nonumber\\
&&\hspace{0.4cm}-\frac{1}{M}\sum\limits_{k=0}^{N}\sqrt{\binom{N}{n+1}\binom{N}{k+1}\binom{N}{k}}[g(\lambda, n+1)!][g(\lambda,k+1)!][g(\lambda,k)!] \nonumber\\
&&\hspace{1cm}\left.\times|Z|^{2n} |Z|^{2k}\sqrt{(n+1)(k+1)}({Z^{2}T^{2}+Z^{\ast2}T^{\ast2}+2|Z|^{2}|T|^{2}})\right\}.
\end{eqnarray}
Similarity, the uncertainty in other quadrature operator ${\hat{Y}}_{R3}$ lead to a similar relation as Eq.~(\ref{uncertainty quadrature X}).
It is clearly seen that the uncertainty in the quadrature operators $\hat{X}_{R3}$ depend on the noise expectation value $\langle F| \hat{F}^{\dag}_{R}\hat{F}_{R}|F\rangle$, and the transmission coefficients $T$.
At finite temperature $\theta$, the noise operators $\hat{F}_R (\omega )$ have the expectation values
\begin{equation}\label{expectation value F}
\langle F| \hat{F}_R^\dagger (\omega ) \hat{F}_R (\omega )| F \rangle =\bar{n}(\omega,\theta)\left( 1-{{\left| R\left( \omega  \right) \right|}^{2}}-{{\left| T\left( \omega  \right) \right|}^{2}} \right),
\end{equation}
where $\bar{n}(\omega,\theta)=1/(\exp[\hbar \omega/k_B\theta]-1)$ is the mean number of thermal photons at frequency $\omega$ in which
$2\pi \hbar$ is Planck's constant and $k_B$ is the Boltzman factor.
Hence, the noise contribution vanishes at zero temperature and as
well for a lossless slab. In this limit cases, we well know that
a CS propagating through slab stays in a minimum uncertainty state
and therefore the output is a coherent light~\cite{Matloob2000}.
Unlike to this situation found for the usual CS, by a simple
checking out of the contents of Eq.~(\ref{uncertainty quadrature
X}), we find that the transmitted nonlinear SCS through the slab
will be no longer a nonlinear CS even in two special cases of
lossless slab and thermally unexcited slab.

Let us study the uncertainty $\langle(\Delta\hat{X}_{R3})^{2}\rangle$ in detail by considering a
single dielectric slab in the ground state and assume a
single-resonance medium of Lorentz type whose complex
permittivity can be given by
\begin{equation}\label{Lorentz model}
{\varepsilon}\left( \omega  \right)=1+\frac{\omega _{P}^{2}}{\omega _{0}^{2}-{{\omega }^{2}}-i\gamma \omega }.
\end{equation}
Here, the parameters ${{\omega }_{P}}$, ${{\omega }_{0}}$ and $\gamma$ refer to the plasma frequency, the resonant frequency and damping coefficient of the dielectric slab, respectively.
Using the equation above, the squares of the absolute values of the calculated reflection, transmission, and absorption coefficients
as functions of frequency ${\omega }/{{{\omega }_{0}}}\;$ are shown in Fig.~\ref{Fig:R2 and T2 versus omega}.
\begin{figure*}[t]
\includegraphics[width=0.5\linewidth]{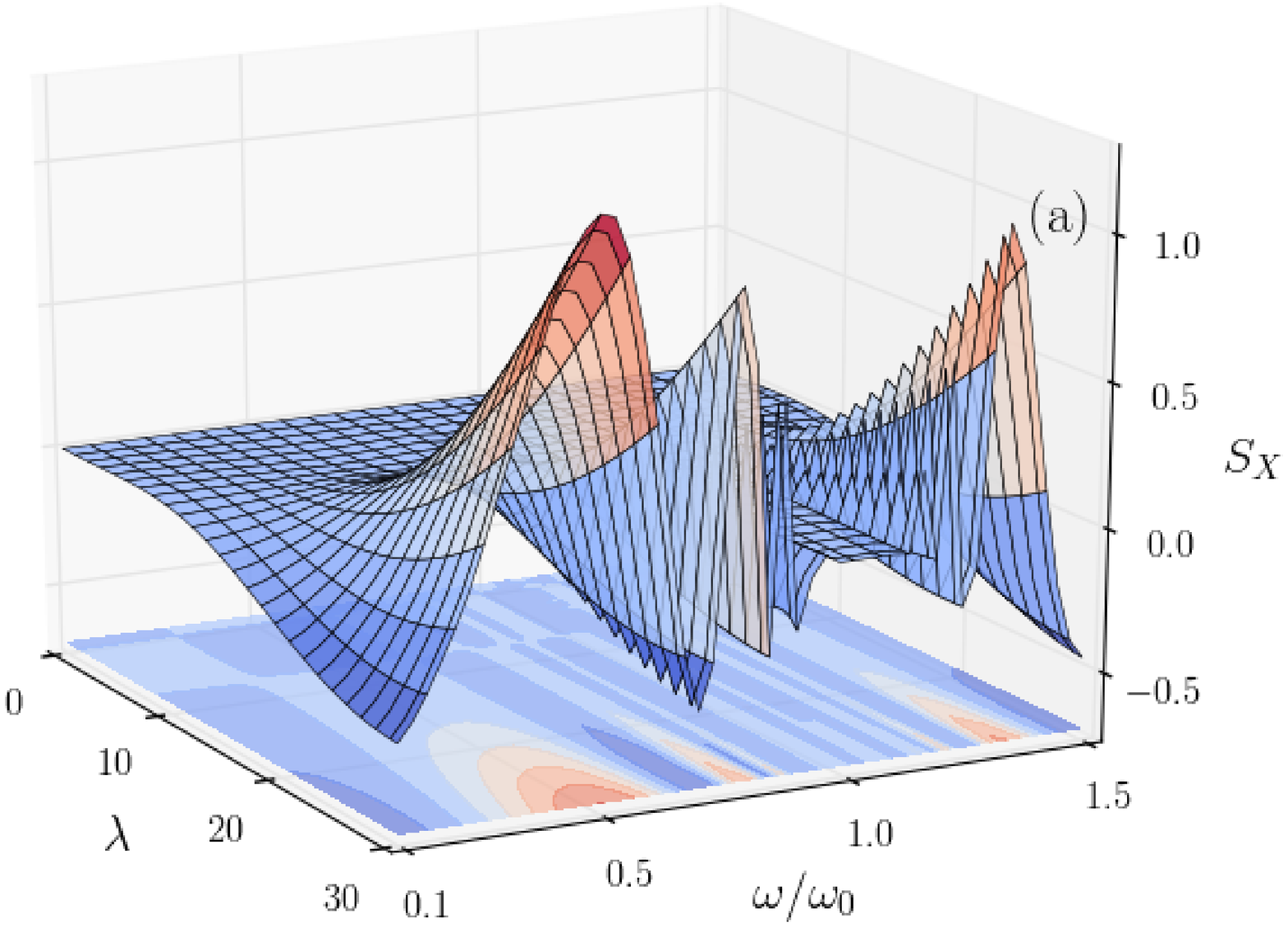}
\includegraphics[width=0.5\linewidth]{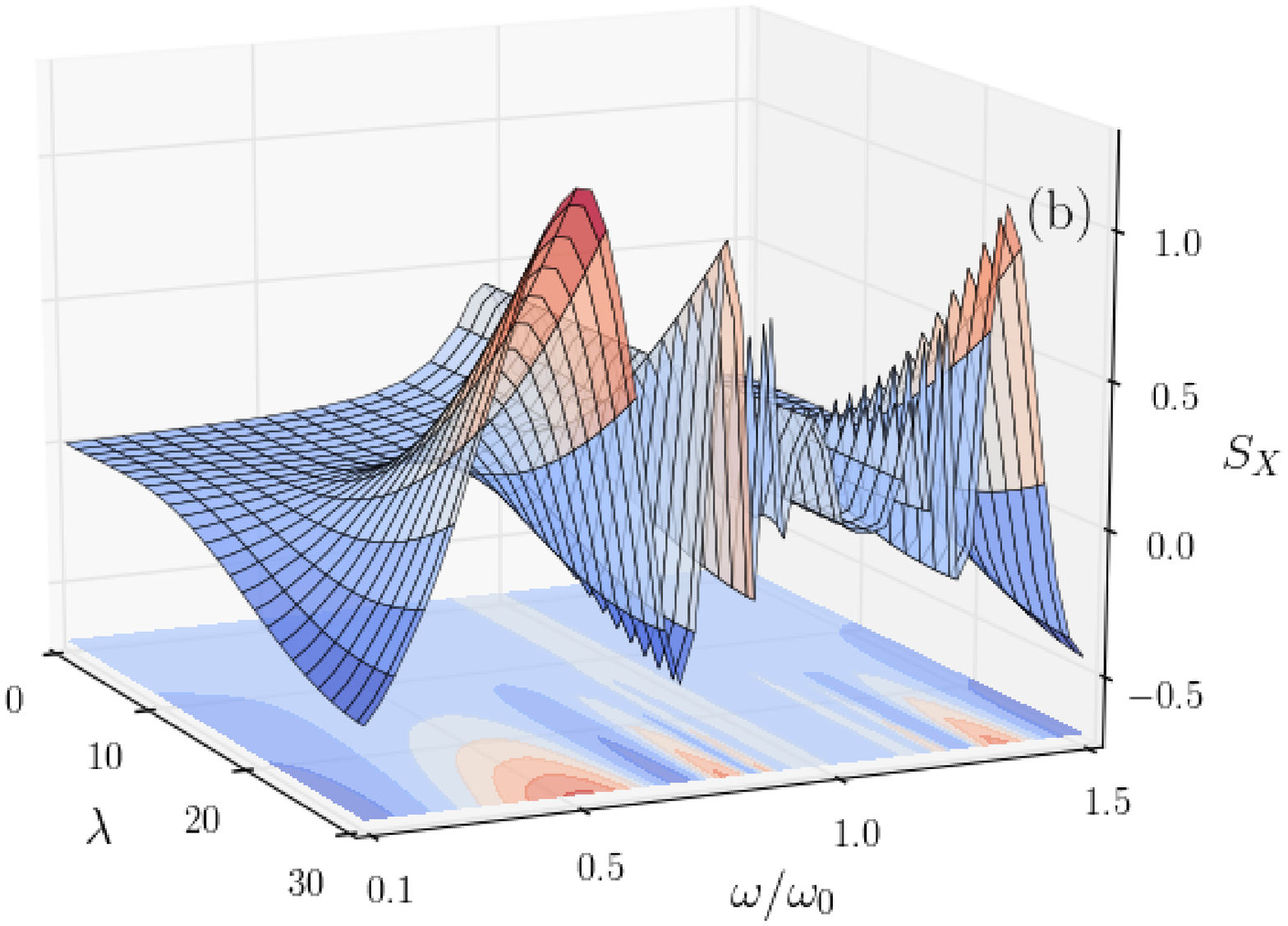}
\caption{Quadrature squeezing variance versus $\lambda $ and ${{{\omega }_{\,}}}/{{{\omega }_{\,0}}}\;$ for a nonlinear coherent state on sphere with $Z=0.01$ and $N=5$ traverses a slab with temperature (a) $\theta=0$ and (b) $\theta={0.6\hbar{{\omega }_{0}}}/{{{k}_{B}}}$. The slab are described by Lorentz model with parameters are identical to those used in Fig.~\ref{Fig:R2 and T2 versus omega}.}
\label{fig:Delta X}
\end{figure*}

For simplicity, we define the parameter ${{S}_{X}}$ as $ {{S}_{X}}=4{{\left( \Delta {{{\hat{X}}}_{R3}} \right)}^{2}}-1$,  where its negative value indicate quadrature squeezing and reduction of quantum noises with respect to standard limit.
In Fig.~\ref{fig:Delta X}, we have plotted ${{S}_{X}}$ versus $\lambda $ and ${\omega }/{{{\omega }_{0}}}$ for different temperatures $\theta =0$ and $\theta=0.6\hbar{\omega }_{0}/k_{B}$. These figures clearly show that squeezing sensitively depends
on the optical properties of the material used for manufacturing
the optical device. Near the resonance frequency, the squeezing parameter ${{S}_{X}}$ is zero at zero temperature, whereas it is always positive at elevated temperatures and its amplitude is increased by increasing the temperature. The reason of this behavior can be explored in Fig.~\ref{Fig:R2 and T2 versus omega}. It is seen that the squares of the module of reflection and transmission coefficient decrease near the
medium resonance. Consequently, the absorption coefficient, $1-{{\left| {{R}}\left( \omega  \right) \right|}^{2}}-{{\left| {{T}}\left( \omega  \right) \right|}^{2}}$, increases in the resonance region.
On one hand, for a slab at finite temperature, the radiated
thermal photons of the medium increases with increasing
temperature. On the other hand, the mean number of thermal
photons and the absorption coefficient appear as a product form
in Eq.~(\ref{expectation value F}). Therefore, we expected that
the squeezing disappears at the elevated temperature near the
resonance frequency. In Fig.~\ref{fig:Delta X} (b), it is also
seen that there is no squeezing at the limit of large physical
curvature. Therefore, the detrimental effect of loss and thermal
noises can not be compensated by increasing the curvature
parameter of the incident state in the resonance region. In other
word, these detrimental effects eliminate the nonclassical
properties of transmitted SCS.

Sufficiently far from the medium resonance, ${\gamma }/{{{\omega }_{\,0}}}\;=0.01$, the relation $|R|^2+|T|^2\approx1$ holds, and the slab acts like a lossless medium, with $|R|^2\approx 0$  and $|T|^2\approx1$.
Therefore, at elevated temperatures, the squeezing is observed in regions where the absorption is weak. It can be shown that the other quadrature component follows a similar behavior.
%
%
\section{Mandel parameter}\label{Sec:Mandel parameter}
%

In order to study the photon number distribution of the
transmitted SCS through the slab at finite temperature, we
calculate the Mandel parameter
\begin{equation}\label{Mandel definition}
Q\equiv \frac{\left\langle {{\left[ \Delta \hat{n} \right]}^{2}} \right\rangle -\left\langle \hat{n} \right\rangle }{\left\langle \hat{n} \right\rangle },
\end{equation}
where the positive, zero and negative values of this parameter refer to super-Poissonian, Poissonian and sub-Poissonian photon-counting statistics, respectively~\cite{Scully1997}.
Recalling Eqs~(\ref{scattering matrix})-(\ref{commutation relation aL1 and aL1dagger}) and make use of the definition of the number operator in region 3, $\hat{n}=\hat{a}_{R3}^\dag \hat{a}_{R3}$, after some algebra, the Mandel parameter for the outgoing state in region 3 can be written as follows,
\begin{eqnarray}
Q&=& \left[ { \sum\limits_{n =0}^{N}\binom{N}{n}[g(\lambda,n)!]^{2}|Z|^{2n}}\right. \nonumber\\
&&\times\left\{  \left({ n^{2}|T|^{4}  -n|T|^{4}+ 4n|T|^{2}\langle F^{\dag}_{R}F_{R}\rangle+2\langle F^{\dag}_{R}F_{R}\rangle^{2} }\right)-\frac{1}{M}\sum\limits_{k =0}^{N}\binom{N}{k} \right.\nonumber\\
&&\times[g(\lambda,k)!]^{2}|Z|^{2k}\left.{ \left. \left({ n k|T|^{4}+\langle F^{\dag}_{R}F_{R}\rangle^{2}+(n+k)|T|^{2}\langle F^{\dag}_{R}F_{R}\rangle }\right) \right\}  }\right]\nonumber\\
&&/\sum\limits_{n =0}^{N}\binom{N}{n}[g(\lambda,n)!]^{2}|Z|^{2n}\{n|T|^{2}+\langle F^{\dag}_{R}F_{R}\rangle\}.
\end{eqnarray}
%
With the help of the Lorentz model~(\ref{Lorentz model}), we show in
Fig.~\ref{fig:Mandel parameter} the effects of loss, the physical curvature and the temperature on the
photon counting statistics by transmitting the sub-Poissonian state ${{\left| Z \right\rangle }_{R}}$ through the dielectric slab at finite temperature.
\begin{figure*}[t]
\includegraphics[width=0.49\linewidth]{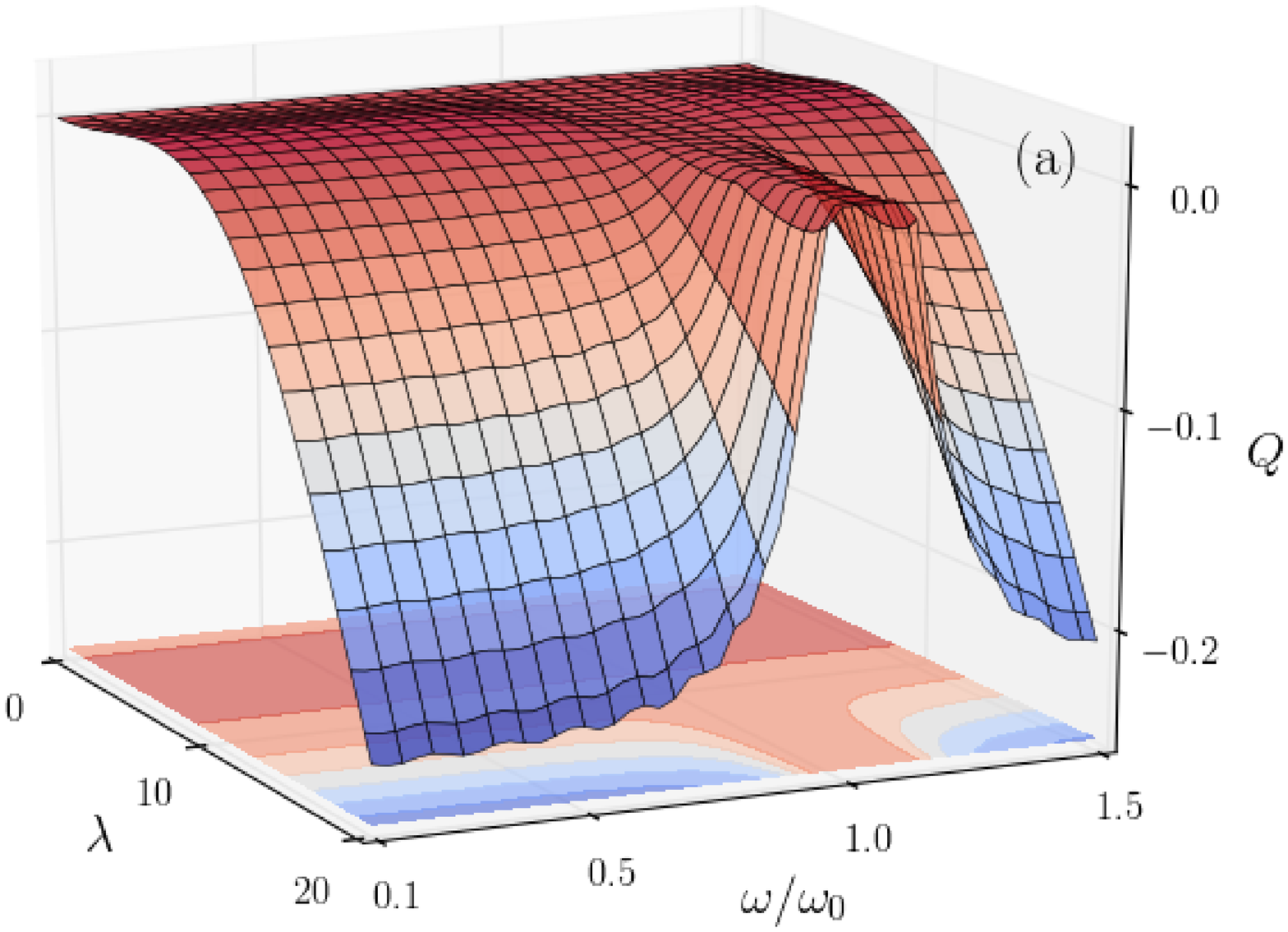}
\includegraphics[width=0.49\linewidth]{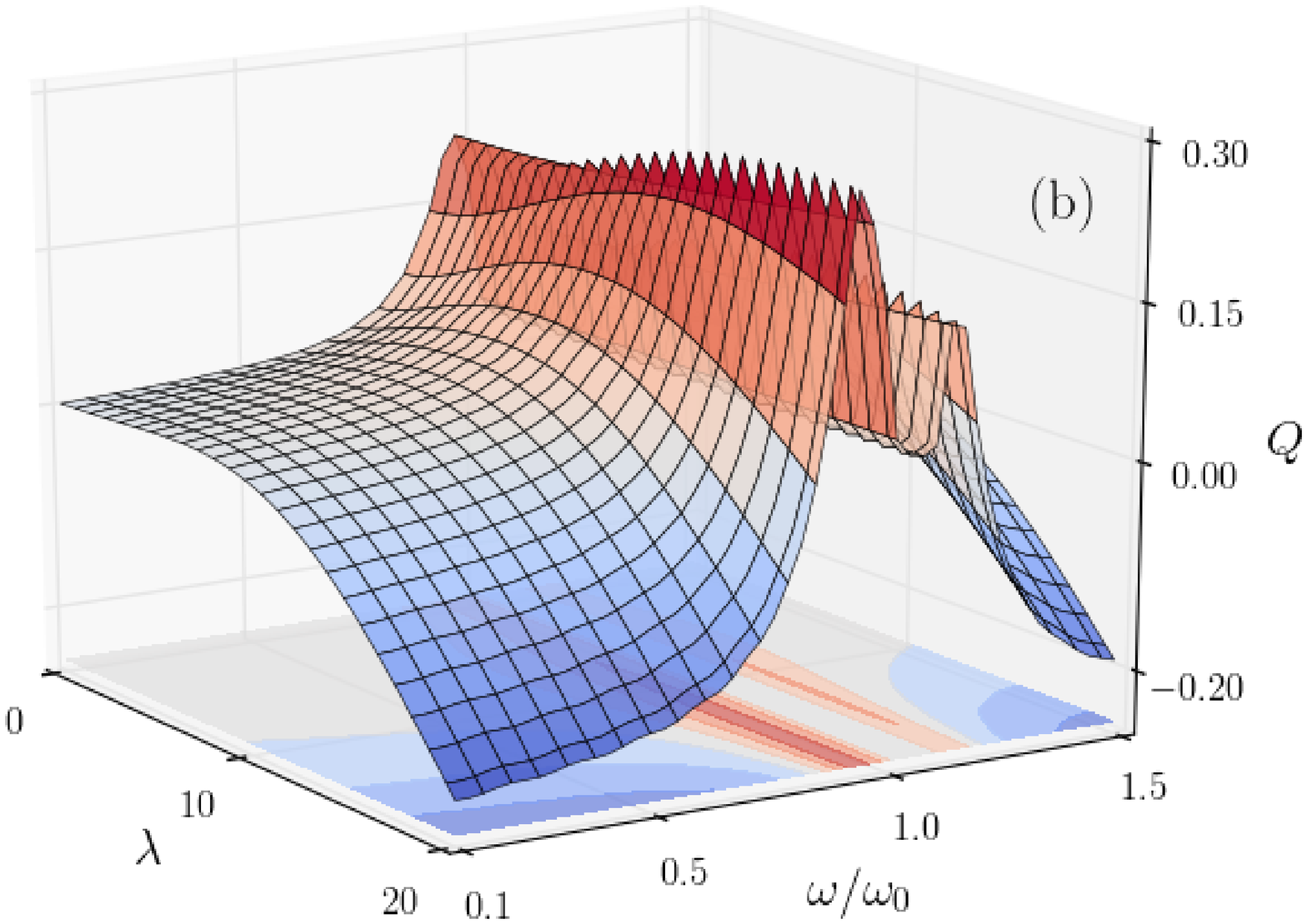}
\centering \caption{Mandel parameter  versus $\lambda $ and
${{{\omega }_{\,}}}/{{{\omega }_{\,0}}}\;$ for a nonlinear SCS
with $Z=0.01$ and $N=5$ traverses a slab with temperature (a)
$\theta=0$, and (b) $\theta={0.6\hbar{{\omega }_{0}}}/{{{k}_{B}}}$. The
slab material parameters are identical to those used in
Fig.~\ref{Fig:R2 and T2 versus omega}.} \label{fig:Mandel
parameter}
\end{figure*}

%
Far from the resonance of medium, the Mandel parameter, independent of the temperature of the slab, is zero and becomes progressively negative with increasing $\lambda$.
As expected, by increasing $\lambda$, the sub-Poissonion nature of outgoing state is somehow amplified.
However, we observe a different behavior near the resonance frequency. At zero temperature, by increasing $\lambda$, the Mandel parameter is likewise zero.
At elevated temperatures, Mandel parameter is positive and become progressively super-Poissonian with increasing temperature. Therefore, as far as it is related to the photon-counting statistics, the dielectric slab in regions where the absorption is strong have the overall effect of degrading the nonclassical
properties of the incident state.

\section{The degree of second-order quantum coherence}\label{Sec:The degree of second order quantum coherence}

In order to investigate how transmission through the slab affects the quantum  coherence properties of incident state, we calculate the degree of the second-order
correlation function~\cite{Scully1997}
\begin{eqnarray}\label{g2 definition}
{{g}^{\left( 2 \right)}}\left( x,t,\tau  \right)=\frac{\left\langle {{{\hat{E}}}^{\left( - \right)}}\left( x,t \right){{{\hat{E}}}^{\left( - \right)}}\left( x,t+\tau  \right){{{\hat{E}}}^{\left( + \right)}}\left( x,t+\tau  \right){{{\hat{E}}}^{\left( + \right)}}\left( x,t \right) \right\rangle }{\left\langle {{{\hat{E}}}^{\left( - \right)}}\left( x,t \right){{{\hat{E}}}^{\left( + \right)}}\left( x,t \right) \right\rangle \left\langle {{{\hat{E}}}^{\left( - \right)}}\left( x,t+\tau  \right){{{\hat{E}}}^{\left( + \right)}}\left( x,t+\tau  \right) \right\rangle },
\end{eqnarray}
for the case that the incident rightwards and leftwards states on
the slab are a continuum nonlinear SCS and the conventional vacuum
state, respectively.
The intensity correlation~(\ref{g2 definition}) determines whether the distribution of the outgoing photons are bunching $({{g}^{\left( 2 \right)}}\left( \tau  \right)<{{g}^{\left( 2 \right)}}\left( 0 \right))$, or antibunching $({{g}^{\left( 2 \right)}}\left( \tau  \right)>{{g}^{\left( 2 \right)}}\left( 0 \right))$.
Here, $\tau$ is the time delay in detecting photons by a coincidence photocount
detector placed at $x$.

For quantum optics problems a model including all modes of the electromagnetic field is required. This is done in two familiar approaches: first a full set of discrete modes in a finite volume of space is used. In second way which is suitable for quantum optical devices with no identifiable cavity, the field is expanded in terms of continuum operators \cite{Blow1990, Barnett}.
Therefore, to consider also these kind of experiments, we generalize the single mode nonlinear SCSs
~(\ref{Nonlinear coherent state}) to the case of continuous
mode nonlinear SCSs as follows:
\begin{eqnarray}\label{continuous nonlinear coherent state}
\hspace{-1cm}\left| \left\{ Z\left( \omega  \right) \right\} \right\rangle &=&\frac{1}{\sqrt{M}}\sum\limits_{m=0}^{N}{\sqrt{\left( \begin{array}{*{20}c}
   N \\
   m \\
\end{array} \right)}\left[ g\left( \lambda ,m \right) \right]!}\frac{{{\left( \int{d\omega \,Z\left( \omega  \right){{a}^{\dagger }}\left( \omega  \right)} \right)}^{m}}}{\sqrt{m!}}\left| 0 \right\rangle,
\end{eqnarray}
where the normalization coefficient, $M$, is given as
\begin{equation}\label{22}
M=\sum\limits_{m=0}^{N}{\left( \begin{array}{*{20}c}
    N \\
    m \\
\end{array} \right){{\left( \left[ g\left( \lambda ,m \right) \right]! \right)}^{2}}{{\left( \int{d\omega {{\left| Z\left( \omega  \right) \right|}^{2}}} \right)}^{m}}},
\end{equation}
and the frequency distribution of the incident light is represented by a function $Z\left( \omega  \right)$ which is determined by the nature of the light
source and any subsequent filtering.
Here, we consider a Gaussian spectrum centered on ${{\omega}_{c}}$ and the mean-square spatial length ${{L}^{2}}$ as
\begin{equation}\label{23}
Z\left( \omega  \right)={{\left( \frac{{{L}^{2}}}{2\pi {{c}^{2}}} \right)}^{{1}/{4}\;}}\exp \left[ {-{{L}^{2}}{{\left( \omega -{{\omega }_{c}} \right)}^{2}}}/{4{{c}^{2}}}\; \right],
\end{equation}
where $c$ is the light velocity. By substituting of Eq.~(\ref{continuous nonlinear coherent state}) in Eq.~(\ref{g2 definition}) and making use of the input-output relation~(\ref{scattering matrix}) and the noise expectation value~(\ref{expectation value F}), after some algebra, we obtain the second-order correlation function in the region 3 as
\begin{eqnarray}\label{g2}
&&g^{(2)}(x, t, \tau)=\frac{1}{M} \left[{ \sum\limits_{n =0}^{N}\binom{N}{n}[g(\lambda,n)!]^{2} \left\{ nJ_{2}(0)|J_{1}(t_{r}+\tau)|^{2}\right.}\right.\nonumber\\
&&+nJ_{2}(0)|J_{1}(t_{r})|^{2}+n(n-1)|J_{1}(t_{r})|^{2}|J_{1}(t_{r}+\tau)|^{2}\nonumber\\
&&\left.{+n(J_{1}^{\ast}(t_{r})J_{1}(t_{r}+\tau)J_{2}^{\ast}(\tau)+c.c.)\}+|J_{2}(\tau)|^{2}+J_{2}(0)^{2} }\right] \nonumber\\
&&/\left\{\left({J_{2}(0)+\frac{1}{M}\sum\limits_{n =0}^{N}\binom{N}{n}n[g(\lambda,n)!]^{2}|J_{1}(t_{r})|^{2}}\right)\right.\nonumber\\
&&\times\left.\left({J_{2}(0)+\frac{1}{M}\sum\limits_{m =0}^{N}\binom{N}{m}m[g(\lambda,m)!]^{2}|J_{1}(t_{r}+\tau)|^{2}}\right)\right\}.
\end{eqnarray}
Here, ${{t}_{r}}=t-x/c$ and the contribution of the transmitted state is described by the integral
\begin{equation}\label{J1}
{{J}_{1}}\left( t \right)=\sqrt{\frac{\hbar }{4\pi {{\varepsilon }_{0}}c \sigma}}\,\int_{0}^{\infty }{d\omega \,}\,{{e}^{-i\omega t}}T(\omega )Z(\omega ),
\end{equation}
and that of the thermal effects of dielectric slab by
\begin{eqnarray}\label{J2}
{{J}_{2}}\left( t \right)&=&\frac{\hbar }{4\pi {{\varepsilon }_{0}}c\sigma }\int{d\omega }\,\omega \,{{e}^{-i\omega t}}\,\bar{n}\left( \omega ,\theta \right) \left( 1-{{\left| {{R}}\left( \omega  \right) \right|}^{2}}-{{\left| {{T}}\left( \omega  \right) \right|}^{2}} \right).
\end{eqnarray}
Because of the complication involved in computing Eq.~(\ref{g2}),
we first investigate some simple limiting cases: In the absence
of the incident nonlinear SCS~(\ref{continuous nonlinear coherent
state}), the degree of second-order coherence~(\ref{g2}) is
simplified to the relation $g_{}^{(2)}(\tau
)=1+\left|\frac{{{ {{J}_{2}}(\tau ) }}}{{{
{{J}_{2}}(0)}}}\right|^2$.
For zero time delay, $g_{0}^{(2)}$ becomes the chaotic value $2$. This is a direct result of the fact that we expected for the thermal light radiated by a thermally excited dissipative slab. While, $g_{0}^{2}$ tends to unity for time delay greater than the thermal coherence time, $\tau_{c}=\frac{\hbar }{{{k}_{\beta }}\theta}$, provided that $\left| {{J}_{2}}(\tau ) \right|$ becomes much smaller than ${{J}_{2}}(0)$.

\begin{figure}[t]
\includegraphics[width=0.5\linewidth]{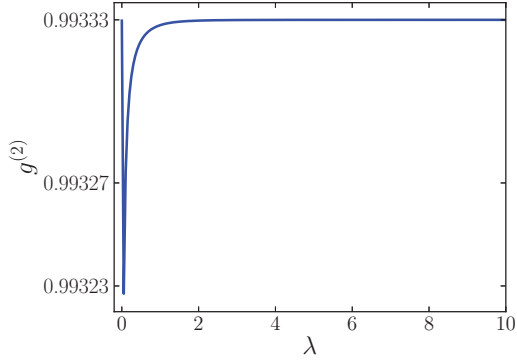}
\centering \caption{The degree of second order quantum coherence
${{g}^{\,(2)}}$ has been plotted versus $\lambda$ for a continuum
nonlinear SCS with $N=50$ traverses through a slab with zero
temperature.} \label{Fig:G2 for T=0}
\end{figure}

In other limiting case, in the presence of incident radiation, we assume that the dielectric slab is kept at $\theta=0$. It is apparent that  ${{g}^{\,(2)}}$ will be independent of the time delay $\tau$. For this case, the second-order correlation function~(\ref{g2}) is plotted as a function of the parameter $\lambda$ in Fig.~(\ref{Fig:G2 for T=0}).
%
%
The results show that $g^{(2)}$ first decreases slightly and then
rises sharply to the same initial value which is nearly unity.
Consequently, the output photons almost independent of the
initial physical curvature exit from the slab with random time
intervals, like the usual CS.

Now, in order to exact investigation of the effects of transmission through the slab and its competition with the curvature of physical space on the bunching and antibunching properties of the incident photons, we need to evaluate the functions ${{J}_{1}}(q)$ and ${{J}_{2}}(q)$.
For the sake of complexity, we first make some simplifications, like what is done in Ref~\cite{Artoni1998b}. We assume that the incident
pulse length is much smaller than the optical thickness of the slab, therefore, the output light appears as a form of a single pulse.
We also take the frequency spread of the continuum nonlinear SCS
to be much smaller than the central frequency.
Hence, with a good approximation, the transmission coefficient $T(\omega )$ in integrals~(\ref{J1}) and~(\ref{J2}) can be replaced by $T(\omega_c)$.
In addition, we assume that the real and imaginary parts of the diffraction index, $\eta(\omega)$ and $\kappa(\omega)$, in the frequency spread of the packet, have constant values which are equal to their values at central frequency ${\omega }_c$, i.e., $\eta_c$ and $\kappa_c$.
Considering above points in mind and making use the relations which was presented in~\cite{Artoni1998b}, the squares of the module of the functions ${{J}_{1}}(q)$ and ${{J}_{2}}(q)$ are given as
\begin{eqnarray}\label{ABS J1 2}
&&|J_{1}(t_{r}+\tau)|^{2}=\frac{\hbar\omega_{c}|T(\omega_{c})|^{2}}{\sqrt{2\pi} \,l\sigma\varepsilon_{0}}\exp[-2c^{2}\tau^{2}]/\\
&&\left\{ L^{2}+2L^{2}(\eta_{c}^{2}-1)^{2}-4L^{3}\kappa_{c}\omega_{c}(\eta_{c}^{2}-1)^{2}(\eta_{c}^{2}+1)/c\eta_{c} \right\},\nonumber
\end{eqnarray}
and
%
\begin{figure*}[t]
\includegraphics[width=0.45\linewidth]{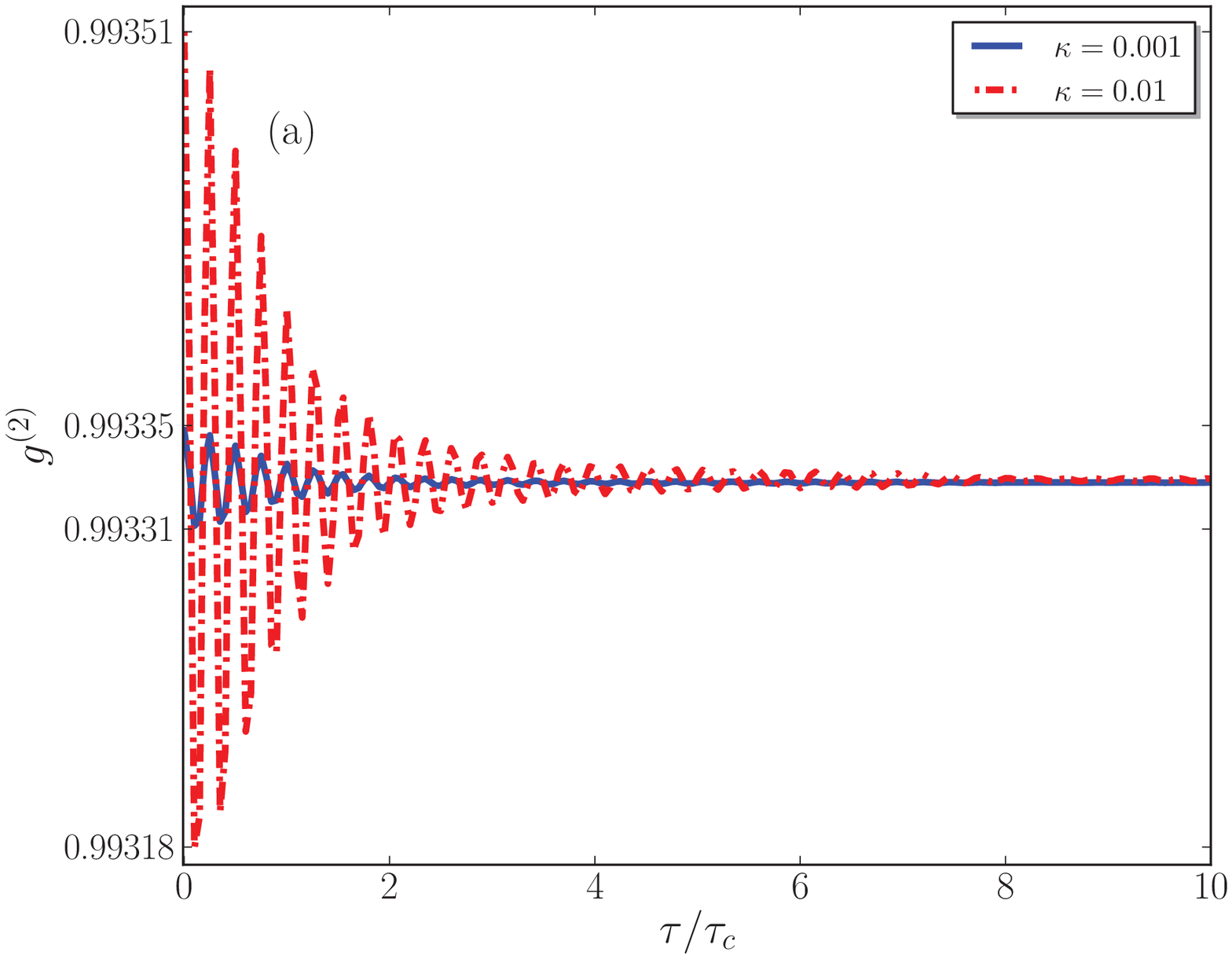}
\includegraphics[width=0.49\linewidth]{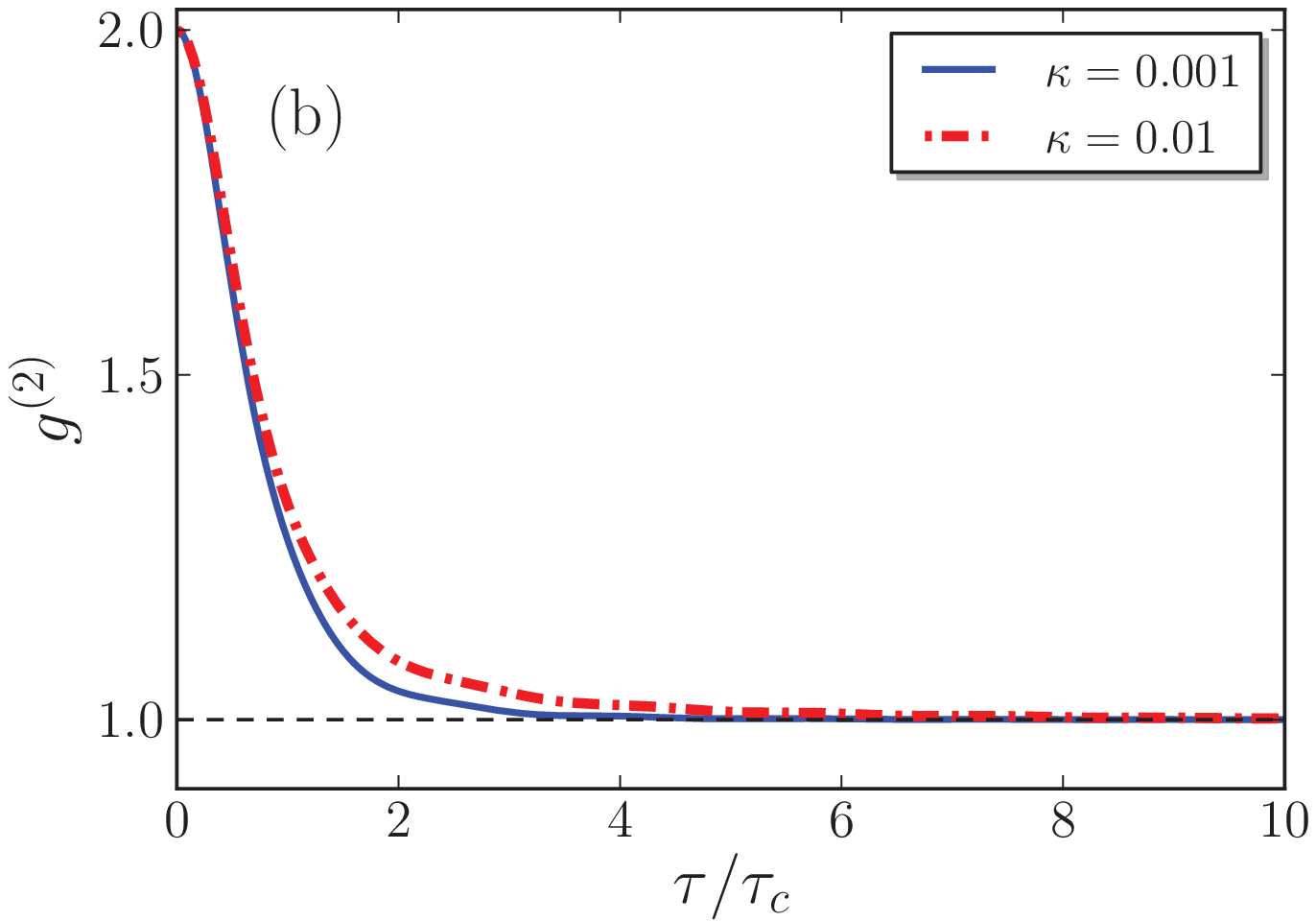}
\centering
\caption{The degree of the second-order correlation function, ${{g}^{\,(2)}}$, versus dimensionless parameter ${\tau }/{{{\tau }_{c}}}$ for a slab with extinction coefficient $\kappa =0.01$ (dash-dotted red curve)  and  $\kappa =0.001$ (solid blue lines) at different temperatures (a) $\theta ={0.0025\hbar {{\omega }_{c}}}/{{{k}_{B}}}$  and (b) $\theta ={0.25\hbar {{\omega }_{c}}}/{{{k}_{B}}}$.  Here, $N=50$ and $\lambda =1$ are chosen and the slab material parameters are identical to those used in Fig.~\ref{Fig:R2 and T2 versus omega}.}
\label{Fig:g2 for T}
\end{figure*}
%
%
%
\begin{eqnarray}\label{ABS J2 2}
&&|J_{2}(t)|^{2}=\frac{\hbar c/\pi \varepsilon_{0}\sigma l^{2}} { [(\eta_{c}+1)^{2}+\kappa_{c}^{2}]^{2}} \left\{ \eta_{c}[(\eta_{c}+1)^{2}+\kappa_{c}^{2}]I_{0}(t,0,0) \right. \nonumber\\
&&-4|n_{c}|^{2}I_{0}(t,4\kappa_{c},0)-\eta_{c}(|n_{c}|^{2}-2\eta_{c}+1)I_{0}(t,8\kappa_{c},0)\nonumber\\
&&+i\kappa_{c}(|n_{c}|^{2}-1)[I_{0}(t,4\kappa_{c},\eta_{c})-I_{0}(t,4\kappa_{c},-\eta_{c})]\nonumber\\
&&\left.+2\kappa_{c}^{2}[I_{0}(t,4\kappa_{c},4\eta_{c})+I_{0}(t,4\kappa_{c},-4\eta_{c})]\right\},
\end{eqnarray}
where
\begin{equation}\label{26}
{{I}_{0}}\left( t,a,b \right)={{\left( \frac{l}{c} \right)}^{2}}\int_{0}^{\infty }{d\omega \,\,\omega \frac{{{\operatorname{e}}^{-\left[ al-i\left( bl+tc \right) \right]{\omega }/{c}\;}}}{{{e}^{{\hbar \omega }/{k_B\theta }\;}}-1}}.
\end{equation}
By inserting the equations above into Eq.~(\ref{g2}), the numerical results of second order quantum correlation degree, ${{g}^{(2)}}$, have been plotted in Fig.~\ref{Fig:g2 for T} as a function of dimensionless parameter $\tau /{{\tau }_{c}}$.
As is seen in Fig.~\ref{Fig:g2 for T} (a), at small temperatures, a damping oscillation behavior in the proximity of unity is observed due to the domination of the pulse contribution. Then, it approaches to a limiting value near unity for time delay greater than $4\tau_c$ and $2\tau_c$, as the incident state passing through the slab with the absorption parameters $\kappa_c=0.01$ and $\kappa_c=0.001$, respectively.
%
%
Here, we consider the parameter $N$ to be 50. Generally, the contribution of higher order terms are negligible compared
to the lower order terms, but if the values of $N>50$ is chosen, the limiting value tends to unity more and more closely, as expected for all large values of $\tau$.
It is worth noting the variation of the curvature parameter $\lambda$ of the incident state, although not shown here, causes no appreciable change in our results.
We therefore find that the transmitted state almost independent of $\lambda$ does not retain its initial antibunched
character at small temperatures.
%

%

In Fig.~\ref{Fig:g2 for T} (b),
$g^{(2)}$ exhibits the chaotic value of 2 near zero time delay as a result of the noise dominates the pulse contribution, then, decrease with $\tau$, and reaching the value of unity over a
time scale of the order of the thermal coherence time. Therefore,  at elevated temperature, we find that the output photons are clearly bunched and this bunching effect enhances slightly by means of increasing loss inside the slab.
%
%
\section{Conclusion}\label{Sec:Conclusion}
%
In this paper, a recently canonical quantization scheme for
layered metamaterials has been used to evaluate the effects of
transmission through a dielectric slab on nonclassical properties
of the nonlinear SCS. To do so, by describing the dielectric
permittivity of the slab by the single-resonance form of the
Lorentz model, some measures of the nonclassicality, including
the squeezing quadrature, the Mandel parameter and the degree of
second-order coherence have been evaluated numerically.

The results show that the non-classical features tend to
survive after propagation through the slab for sufficiently far from the medium resonance, even their magnitudes were enhanced by increasing the physical space curvature of the incident state.
While, the effects of transmission through the slab are more
drastic in the resonance region, and antibunching, sub-Poissonian
statistics and squeezing effects are usually lost due to the loss
within the slab and as well the thermal noise inevitably
accompanies the propagation of the input photons through the slab
at elevated temperatures. Thus, for a single and continuum SCSs
considered here, respectively, both quadrature squeezing and
Mandel parameter, and the degree of second-order coherence tend
towards classical values after transmission, and this behavior is
enhanced when the slab is kept at an elevated temperature.
Finally, we found that the detrimental effects of the loss and
the thermal noise, near the medium resonance, can not be
compensate by increasing the physical space curvature of the
incident state.
%
\section*{Acknowledgments.}
Authors wish to thank the Shahrekord University for support.
%
%

\end{document}